\documentstyle[12pt]{article}
\newcommand{\al}{\alpha}
\newcommand{\si}{\sigma}
\newcommand{\om}{\omega}
\newcommand{\fee}{\varphi}

\newcommand{\pa}{\partial}
\newcommand{\be}{\begin{equation}}
\newenvironment{punto}[1]{\begin{itemize}{\item\bf #1\/}:}{\end{itemize}}
\newcommand{\ee}{\end{equation}}
\newcommand{\bea}{\begin{eqnarray}}
\newcommand{\eea}{\end{eqnarray}}
\begin{document}
\begin{center}
{\LARGE {\bf Composite Weak Bosons in a Confining Gauge Theory
without Goldstone Bosons: a Strong Coupling Expansion Analysis}}\\[0.9cm]
{\Large A. Galli\\[0.3cm]
{\em Paul Scherrer Institute, CH-5232 Villigen PSI, Switzerland}}\\[0.3cm]
{\em \today}\\[1cm]
{\Large {\bf Abstract}}
\end{center}
We consider a confining Yang-Mills theory without Goldstone Bosons which could
describe the bosonic sector of the weak interactions.
This model can be gauge invariantly regularized on a
lattice. A strong coupling analysis of the low lying bound state spectrum
indicates that the vector isotriplet bound state (the right quantum number to
represent the W-boson) could be the lightest
state if the mass of the pseudoscalar isosinglet is raised sufficiently by the
effect of the chiral anomaly (in analogy to the $\eta'$ mechanism of QCD).\\
This work is a preliminary study in support to an intensive lattice Monte Carlo
analysis of the model.
\newpage
\section{Introduction}

Today,
all the known fundamental interactions of elementary particles (strong, weak
and electromagnetic) are described by the Standard Model (SM) \cite{SM}.
The matter
fields of the SM (quarks and leptons) and the Higgs boson
are considered to be elementary. They
interact with each other by the exchange of gauge bosons which are also
considered to be elementary. The structure of the theory is essentially
determined once the matter fields and their transformation laws under the local
gauge transformations are specified.
The SM has received a great deal of
phenomenological support and the structure of the model has been confirmed to
a high degree of accuracy.\\
In spite of the beautiful corroboration of the SM by experiments, we may
naturally ask the questions: how elementary are quarks and leptons, the gauge
fields and the Higgs boson? The possibility of discovering further
substructure within the particles of the SM remains a viable option for the
physics which lies beyond that model.\\
The aim of introducing substructures is to construct a simple
fundamental theory with a few degrees of freedom which
should be able to reproduce the SM as an effective theory.
In this way one hopes to be able to solve three important theoretical problems
of the
SM: 1) the family replication of the matter fields, 2) the existence of too
many parameters and 3) the fine tuning problem.
Several models treat the quarks and lepton, the gauge bosons and the Higgs
boson as composite systems. Today, a conspicuous number of theorems exists
which
have ruled most of the existing models \cite{2} and radically
restricted the possibilities to construct realistic composite models.
Two models have survived: the "Strongly coupled SM" (SCSM) \cite{SCSM} and
the Yang-Mills theories without Goldstone bosons \cite{SH}. The aim of the SCSM
is
not to construct a simple fundamental theory beyond the SM, but to propose an
alternative to the usual Higgs mechanism. However, the Yang-Mills theory
without Goldstone bosons proposes to describe composite weak bosons starting
from simple fundamental principles.\\
To be precise this last model considers the photon to
remain elementary and switched off. The weak gauge bosons $W^\pm$ and $Z$ then
form a mass degenerate triplet.
This model is a usual
Yang-Mills theory, with gauge fields and fermion fields, but with a special
choice of the degrees of freedom. The degrees of freedom are characterized
by the local gauge group, by the global isospin group and by the choice of
the fermion fields (Dirac or Majorana spinors). To be plausible
a composite model of the
weak bosons has to reproduce the known weak boson spectrum: the lightest
bound states have to be the W-bosons and heavier bound states have to lie
in an experimentally unexplored energy range.
The only possibility to have a Yang-Mills theory which reproduces the weak
boson spectrum is to choose the degrees of freedom in a way
that they naturally avoid bound states lighter than the vector isotriplet
of the theory (which characterizes the W-boson triplet).
This is possible if the
unwanted light bound states which naturally show up as Goldstone bosons or
pseudo Goldstone bosons in many models
(like, for example, a pseudoscalar iso-nonsinglet,
which would be the pion analogue of QCD) vanish by the Pauli principle, i.e.
they are symmetric combinations of Grassmann variables.
To obtain such a theory it is sufficient to choose the hypercolor
gauge group to be
$SU(2)$, the isospin group to be $SU(2)$ and the fermion fields to be
Majorana spinors.\\

In our work we will investigate the spectrum of this theory for two
reasons: first, it is a very interesting project by itself to investigate the
spectrum of a confining theory which exhibits a spectrum which is completely
different from the QCD state spectrum and, second, this theory
contains fundamental ingredients needed to describe a realistic model of
composite weak bosons.\\
Because of the non-perturbative nature of the problem the model must be treated
on a lattice in Euclidean space. We emphasize that
the spectrum in Euclidean space is equal to the spectrum in Minkowski
space. A lattice regularization \`a la Wilson \cite{WL} is possible because the
choice
of the isospin group $SU(2)$ allows us to replace the Dirac mass term and the
Dirac-type Wilson term by a hypercolor gauge invariant Majorana expression.\\

As a first step of our investigation we present a strong coupling expansion.
We will explain in a formal way the technic that we have introduced
to perform the strong coupling expansion.
In particular we have developed an
algorithmic way to easily identify and evaluate
the diagrams of the expansion. This algorithm is very useful because
the mass splitting between the lightest bound states show up only at
higher order of the strong coupling expansion and to obtain a significant
result
one has to evaluate many hundreds of diagrams.\\
{}From the result of this calculation we show that the vector isotriplet bound
state of this theory is lighter than all other bound states provided
that the chiral anomaly rises sufficiently the pseudoscalar isosinglet mass, in
analogy to the $\eta'$ mechanism in QCD \cite{ETA}. \\
This work is a preliminary analysis in support to an intensive lattice Monte
Carlo simulation where the mass of the pseudoscalar isotriplet is evaluated
explicitely \cite{galli}.
Outline of the paper: In section 2 we describe the model, in section 3 we
explain the technic of our strong coupling expansion, in section 4 we classify
the composite operators according to the CP eigenvalues and in section 5 we
evaluate the spectrum of these CP eigenstates.

\section{The Confining Gauge Theory without Goldstone Bosons}
We start with a definition of our model\footnote{In \cite{SH} an unfortunate
sign error has entered the calculations due to a wrong factor $i$ in the fourth
and fifth term of eq. (3). To interpret the result correctly one has to
interchange isosinglet with isotriplet. I thank H.Schleret for a discussion of
this point.}
\cite{SH}.
For pedagogical reasons,
in this section, we formulate our theory using Weyl spinors
in a second step constructing the Majorana spinors and rewriting the action in
a QCD-like form.\\
\subsection{Definition of the fermion fields}
We consider a gauge theory whose fermion content is represented by a Weyl
spinor
$F_{\al,a}^A(x)$. Here $\al$ denotes the (undotted) spinor index ($\al=1,2$),
$A$
denotes the fundamental representation index of a global SU(2) isospin group
($A=1,2$) and $a$ denotes the fundamental representation index of the local
SU(2) hypercolor gauge group ($a=1,2$). Starting
from this fermion field F the following
hypercolor singlet operators for the low lying bound states can be formed:
\begin{punto}{Lorentz scalars} with fermion number +2 resp. -2
\bea
S(x) &=& F(x)QF(x)\nonumber\\
\bar S(x) &=& F^\dagger (x) QF^\dagger (x)
\eea
\end{punto}
\begin{punto}{Lorentz vectors} isosinglet resp. isotriplet
\bea
V_\mu(x)&=& F^\dagger(x)\si_\mu F(x)\nonumber\\
V_\mu^I(x) &=& F^\dagger(x)\si_\mu T^I F(x)
\eea
\end{punto}
\begin{punto}{Lorentz tensor}
\bea
B_{\mu\nu}^I(x)&=&F(x)Q\tilde\si_\mu\si_\nu T^IF(x)\nonumber\\
\bar B_{\mu\nu}^I(x)&=&F^\dagger(x)\si_\mu\tilde \si_\nu T^IQF^\dagger(x)
\eea
\end{punto}
where $\{\si_\mu\}=(i,\si_j)$ ($\mu=0,\dots,3$)
denotes the four component generalization of the
Pauli matrices acting on the spin indices of the Weyl spinors
in the Euclidean space-time and
$\{T^I\}$ ($I=1,2,3$) are the Pauli matrices of the global isospin group.
The matrix Q represents the antisymmetric matrix in spin, hypercolor and
isospin space, which correspond to the Kronecker product of
$i\si_2,i\tau_2,iT_2$ (the antisymmetric matrices in spin, hypercolor and
isospin space, respectively)
$$
Q=i\si_2\otimes i\tau_2\otimes iT_2
$$
The notation F(x)QF(x), e.g., is a shorthand for the contraction of
all the spin, isospin and hypercolor indices of the fermion fields with the
matrix Q.
Obviously among the composite fields (1), (2) and (3) there is
only one candidate for a Goldstone boson, namely S,
associated with a spontaneous breakdown of the
fermion number transformation group
$F\rightarrow e^{i\al}F$ by the condensate
$\langle\bar S\rangle=\langle S\rangle\neq 0$,
which should arise in the confining phase. It is
expected that the CP odd part of S (namely, $S-\bar S)$ acquires a mass by the
chiral
anomaly in analogy to the $\eta'$ in QCD. Notice that a
Lorentz scalar isotriplet operator vanishes by the Pauli principle.
As discussed above this last property is fundamental for any realistic
electroweak composite model.
\subsection{The Wilson lattice action}
There is a SU(2) hypercolor matrix U(b) (in the fundamental representation)
defined
on each oriented lattice bond b. Our convention is that
\be
U(-b)=U^\dagger(b)
\ee
An oriented path $\om$ on the lattice is a set of bonds
\be
\om=b_1\cup b_2\cup\dots\cup b_n
\ee
such that the endpoint of $b_i$ is the startpoint of $b_{i+1}$ for
$1\leq i\leq n-1$. We can associate a SU(2) hypercolor matrix with $\om$ by
defining
the path ordered product
\be
U(\om)=U(b_1)U(b_2)\dots U(b_n)
\ee
The $\si-$matrices are defined as follows:
\be
\si(b)=\left\{\begin{array}{cc}
\si_\mu &\mbox{ if b in $+\mu$ direction}\\
-\si_\mu &\mbox{ if b in $-\mu$ direction}
\end{array}\right.
\ee
We will denote by $p\subset \Lambda$ an elementary
plaquette of the lattice $\Lambda$ and with $\pa p$ the boundary of $p$.
The complete Wilson fermion action on a lattice $\Lambda$
written in the Weyl spinor notation is given by

\bea
S&=&\beta\sum_{p\subset \Lambda}TrU(\pa p)\nonumber\\
& &-2k\sum_{b=\langle xy\rangle\subset\Lambda}\left\{F^\dagger(x)\si(b)U(b)F(y)
+\frac{r}{2}\left(F(x)QU(b)F(y)+F^\dagger(x)U(b)QF^\dagger(y)\right)\right\}\nonumber\\
&
&-\frac{1}{2}\sum_{x\in\Lambda}\left\{F(x)QF(x)+F^\dagger(x)QF^\dagger(x)\right\}
\eea
The parameter $\beta$ is related to
the bare coupling constant by $\beta=\frac{4}{g^2}$, k is the hopping
parameter and r (=1) multiplies the Majorana version of the Wilson term. The
Weyl
fermions are represented by the Grassmann variables $F(x)$ and $F^\dagger(x)$.
Lattice points are denotes by $x\in\Lambda$. $\langle xy\rangle\subset\Lambda$
represent the
oriented bond between nearest neighbour pairs of lattice points.

\subsection{Compact notation}
Defining $\tilde\si^\mu=(-i,\si^j)$ one can easily check the following three
relations
\bea
i\tau_2U(b)^\dagger i\tau_2&=&-U(b)^T\\
i\si_2\tilde\si^\mu i\si_2&=&\si^{\mu T}\\
QU(b)^\dagger\tilde\si^\mu Q&=&U(b)^T\si^{\mu T}
\eea
They can be used to rewrite the first part of the fermionic kinetic term
in the action as a sum of two equivalent parts
\bea
& &\sum_{b=\langle xy\rangle}F^\dagger(x)\si(b)U(b)F(y)=\nonumber\\
& &\frac{1}{2}\sum_{b=\langle xy\rangle}F^\dagger(x)\si(b)U(b)F(y)+
\frac{1}{2}\sum_{b=\langle xy\rangle}F(x)Q\tilde\si(b)U(b)QF^\dagger(y)
\eea
This observation allows us to define the following four-spinor fermion
field and to rewrite the action in a compact form, which is similar to the
Wilson action of QCD:
\be
\fee^A_{\tilde\al a}(x)=\left(\begin{array}{c}
F_{\al,a}^A(x)\\F^{\dagger^T A}_{\dot\al,a}(x)
\end{array}\right)
\ee
where the index $\tilde\al$ runs from 1 to 4 and indicates the two upper spinor
components $F_{1,a}^A$ and $F_{2,a}^A$ for $\tilde\al=1,2$ and the two lower
spinor components $F^{\dagger A}_{\dot 1,a}$ and $F^{\dagger A}_{\dot 2,a}$ for
$\tilde \al=3,4$.
Its transpose is simply defined by $\fee^T=(F^T,F^\dagger)$.\\
In the notation (13) the
isospin transformation $F\rightarrow e^{i\vec{\al}\vec{T}}F$
is
$$
\fee\rightarrow exp\left\{i\vec{\al}
\left(\begin{array}{cc}\vec{T}&0\\0&Q\vec{T}Q^T\end{array}\right)\right\}\fee
$$
The spin matrices are defined in terms of ordinary $\gamma$-matrices by
\be
\Gamma(b)=\left\{\begin{array}{cc}
\Gamma^\mu=r+\gamma^\mu&\mbox{if b in $+\mu$ direction}\\
\bar\Gamma^\mu=r-\gamma^\mu&\mbox{if b in $-\mu$ direction}
\end{array}\right.
\ee
The $\gamma$-matrices are hermitian 4x4 matrices, satisfying
$\{\gamma^\mu,\gamma^\nu\}=2\delta^{\mu\nu}$ which we choose
as follows:
\be
\gamma^\mu=\left(\begin{array}{cc}
0&\tilde\si^\mu\\\si^\mu&0
\end{array}\right)
\ee
and we also define
\be
\gamma_5=\gamma_0\gamma_1\gamma_2\gamma_3= \left(\begin{array}{rr}
-1&0\\0&1
\end{array}\right)
\ee
With this choice of $\fee$ and $\Gamma(b)$ and with eq. (12)
we can rewrite the action (8) in the
following compact form
\bea
S&=&\beta\sum_{p\subset \Lambda}TrU(\pa p)\nonumber\\
& &-k\sum_{b=\langle xy\rangle}\fee^T(x)
\left(\begin{array}{cc}
Q&0\\0&1\end{array}\right)
\Gamma(b)U(b)
\left(\begin{array}{cc}
1&0\\0&Q\end{array}\right)
\fee(y)\nonumber\\
& &-\frac{1}{2}\sum_{x}\fee^T(x)A\fee(x)
\eea
where the matrix A is defined by
\be
A=\left(\begin{array}{cc}
Q&0\\0&1\end{array}\right)
\left(\begin{array}{cc}
1&0\\0&Q\end{array}\right)=
\left(\begin{array}{cc}
Q&0\\0&Q\end{array}\right)
\ee
We note that the action (17) is invariant under the isospin transformation.
In section 4 we will go one step further and use
fermion fields of the form
\be
\psi=\left(\begin{array}{cc}1&0\\0&Q\end{array}\right)\fee\,\,\,
\mbox{and}\,\,\,\bar\psi=\fee^T
\left(\begin{array}{cc}Q&0\\0&1\end{array}\right)
\ee
in terms of which the action closely resembles the Wilson action of QCD.
\bea
S&=&\beta\sum_{p\subset\Lambda}TrU(\pa p)-k\sum_{b=\langle
xy\rangle}\bar\psi(x)
\Gamma(b)U(b)\psi(y)-\frac{1}{2}\sum_{x}\bar\psi(x)\psi(x)\equiv\nonumber\\
&\equiv&S_{Gauge}(U)-\bar{\psi} \frac{M(U)}{2} \psi\nonumber
\eea
We emphasize that one has to be careful since, unlike in QCD, $\bar \psi$
and $\psi$ are not independent fields.
They are Majorana fields with hypercolor and isospin.

\section{Strong coupling expansion rules}
We will use functional techniques to give a simple derivation of Wilson's
Feynman rules for the strong coupling expansion that we will formulate on the
lattice. By coupling the various degrees of freedom to sources, we will obtain
a
compact formal expression for the Green's functions of the theory. Strong
coupling expansions have been discussed in [7,8]. Here we analyse the changes
needed for the Majorana case for an expansion requiring high order.

\subsection{Functional integration}
Quantization of the fermion degrees of freedom can be carried out by expressing
the
transition amplitude directly as a sum over all possible paths in field space
connecting the initial and the final states. The partition sum is then
$$
Z=\int[d\fee][dU]exp\left\{-S(\fee,U)\right\}
$$
The sum over the paths for the boson degrees of freedom is a functional
integral over the
element of the gauge group U(b):
\be
[dU]=\prod_{b\subset \Lambda}dU(b)
\ee
The sum over the paths for the fermion degrees of freedom is a functional
integral over the
anticommuting fermion fields $\fee(x)$:
\be
[d\fee]=\prod_{x\in\Lambda}d\fee(x)
\ee
Thus they are elements of a Grassmann algebra. Notice that in contrast to QCD
here there are no independent conjugate variables $\bar\fee$. The integration
(21) is defined as follows

\begin{punto}{Grassmann integration} In a n-dimensional Grassmann algebra,
the n Grassmann generators
$\theta_1,\theta_2,\dots,\theta_n$ satisfy $\{\theta_i,\theta_j\}=0$ (for
$i,j=1,2,\dots,n$). The symbols $d\theta_1,d\theta_2,\dots,d\theta_n$ are
defined by their algebraic properties $\{d\theta_i,d\theta_j\}=
\{d\theta_i,\theta_j\}=0$ and
\bea
\int d\theta_i&=&0\nonumber\\
\int d\theta_i\theta_j&=&\delta_{ij}\nonumber
\eea
{}From these rules one derives that for any {\em totally antisymmetric} matrix
$A$
\be
\int d\theta_1d\theta_2\dots d\theta_n exp(\frac{1}{2}(\theta,A\theta))=
\sqrt{det\,A}
\ee
which is the essence of the Grassmann integration. We obtain also a formula
with shifted variables
\be
\int d\theta_1d\theta_2\dots d\theta_n
exp(\frac{1}{2}(\theta,A\theta)+(\eta,\theta))=
\sqrt{det\,A}\,\,\,\exp(-\frac{1}{2}(\eta,A^{-1}\eta))
\ee
where $\eta=(\eta_1,...,\eta_n)$ are Grassmann variables.
\end{punto}
The strong coupling expansion is most easily derived and compactly
formulated through the introduction of external sources coupled to the degrees
of freedom of the theory. In this approach, Green's functions are obtained by
differentiating with respect to the various sources. Corresponding to the
degrees of freedom $U^{aa'}(b)$ and $\fee_{\tilde\al,a}^A(x)$ we introduce the
sources $K^{aa'}(b)$ and $\lambda_{\tilde\al,a}^A(x)$,
respectively\footnote{
$\lambda(x)$ can be expressed by two sources associated to the
two Weyl fermion fields $F(x)$ and $F^\dagger(x)$,
respectively:
$\lambda^A_{\tilde \al,a}=\left(\begin{array}{c}
\eta_{\al,a}^A(x)\\\bar\eta^{A}_{\dot\al,a}(x)
\end{array}\right)$. Notice that the bar over $\eta$ simply denotes that $\eta$
and $\bar \eta$ are two different sources.}.

Perturbation theory begins by breaking the action
S into two parts: the unperturbed action
$S_0$ and the perturbation action $S_I$. For our strong coupling expansion we
choose
\bea
S_0&=&-\frac{1}{2}\sum_{x}\fee^T(x)A\fee(x)\\
S_I&=&\beta\sum_{p}TrU(\pa p)-k\sum_{b=\langle xy\rangle}\fee^T(x)
\left(\begin{array}{cc}
Q&0\\0&1\end{array}\right)
\Gamma(b)U(b)
\left(\begin{array}{cc}
1&0\\0&Q\end{array}\right)
\fee(y)
\eea
A n-points Green's function can then be written as
\bea
& &\langle\fee^{A_1}_{\tilde\al_1a_1}(x_1)\dots
\fee^{A_n}_{\tilde\al_na_n}(x_n)\rangle=\nonumber\\
& &=\frac{1}{Z}
\int[d\fee][dU]\fee^{A_1}_{\tilde\al_1a_1}(x_1)\dots
\fee^{A_n}_{\tilde\al_na_n}(x_n)exp\{-S(\fee,U)\}=\\
& &=\frac{1}{Z}\;
\frac{\delta}{\delta \lambda^{A_1}_{\tilde\al_1a_1}(x_1)}\dots
\frac{\delta}{\delta \lambda^{A_n}_{\tilde\al_na_n}(x_n)}
exp\left\{-S_I\left(\frac{\delta}{\delta \lambda},\frac{\delta}{\delta
K}\right)\right\}
\times\nonumber\\
& &\times
\left.\int[d\fee][dU]exp\left\{-S_0(\fee,U)+\sum_{x}\lambda^T(x)\fee(x)+
\sum_{b}K(b)U(b)\right\}\right|_{\lambda,K=0}\nonumber
\eea
where $Z=\int[d\fee][dU]exp\{-S(\fee,U)\}$.
The notation $\lambda^T\fee$, for example, is a shorthand for the contraction
of
the hypercolor, spin and isospin indices. The division by Z is equivalent to
the
omission of disconnected diagrams with at least one factor being a vacuum
diagram (no external legs).\\
We note that the unperturbed action depends on the fermion fields only.
The functional integration over the fermion fields yields (using eq. (22-23))
\bea
& &\int[d\fee]exp\left\{-S_0(\fee)+\sum_{x}\lambda^T(x)\fee(x)\right\}=
\nonumber\\
& &=\int[d\fee]exp\left\{\sum_{x}\left[\frac{1}{2}\fee^T(x)A\fee(x)+
\lambda^T(x)\fee(x)\right]\right\}=\nonumber\\
& &=exp\left\{-\sum_{x}\lambda^T(x)\frac{A^{-1}}{2}\lambda(x)\right\}
\eea
The functional integral over the gauge group gives us
\be
\int[dU]exp\left\{\sum_{b}K^{aa'}(b)U^{aa'}(b)\right\}=\prod_{b}D(K(b),K(-b))
\ee
where $\prod_{b=\langle xy\rangle}$ denotes a product over all nearest
neighbour pairs of lattice points and
\be
D(K(b),K(-b))=\int
dg\,\,exp\left\{K^{aa'}(b)g^{aa'}+K^{aa'}(-b)(g^{-1})^{aa'}\right\}
\ee
Here $dg$ denotes the Haar measure over the gauge group.
Putting all this together, we obtain the expression for the n-point Green's
function (we omit the hypercolor, isospin and spin indices for simplicity)
\bea
\langle\fee(x_1)\dots\fee(x_n)\rangle
&=&\frac{1}{Z}\;
\frac{\delta}{\delta \lambda(x_1)}\dots\frac{\delta}{\delta \lambda(x_n)}
exp\left\{-S_I\left(\frac{\delta}{\delta \lambda},\frac{\delta}{\delta
K}\right)\right\}
\times\nonumber\\
& &\times \left.
exp\left\{-\sum_{x}\lambda^T(x)\frac{A^{-1}}{2}\lambda(x)\right\}\times
D(K(b),K(-b))\right|_{\lambda,K=0}\nonumber\\
& &=\left.\frac{\delta}{\delta \lambda(x_1)}\dots\frac{\delta}{\delta
\lambda(x_n)}
exp\left\{-S_I\left(\frac{\delta}{\delta \lambda},\frac{\delta}{\delta
K}\right)\right\}\times \tilde Z[\lambda,K]\right|_{\lambda ,K=0}
\eea
where the last factor $\tilde Z$ denotes the generating functional
\be
\tilde Z[\lambda,K]=\frac{1}{Z}\;
exp\left\{-\sum_{x}\lambda^T(x)\frac{A^{-1}}{2}\lambda(x)\right\}\times
\prod_{b}D(K(b),K(-b))
\ee
The expansion of the exponential of the perturbation part of the
action $exp\{-S_I\}$ in a power
series generates the strong coupling series. The terms of this series may be
represented by graphs similar to the Feynman graphs in ordinary perturbation
theory.
The graphical rules for calculating
the Green's function can be read off from the equation (30). In the next
subsections we will present simple algorithmic Feynman rules for the two point
Green's function of the form
$\langle \fee(x)H\fee(y)\rangle$ or
$\langle \fee(x)M_1\fee(x)\fee(y)M_2\fee(y)\rangle$ where
$H$, $M_1$ and $M_2$ denote
some general matrices in spin, isospin and hypercolor space which produce
hypercolor singlets and respect the
covariance of the bilinear forms $\fee M_j \fee$ (for example $\fee\fee$ is not
covariant\footnote{One can't contract dotted spinor indices with undotted
ones.}).
For the fermion string
through the graphs along the path $\om$ we only need the familiar
Dirac-fermion rules. But for the
vertices at the external points x and y the rules are different from the
Dirac-fermion ones, due to the Majorana nature of the fermion.

\subsection{Group integrals}
In equation (30) we are left with the integral $D(K(b),K(-b))$. This integral
is a generating functional for integrals of polynomials of group matrices
trough
the relation
\bea
& &\int dg g^{a_1a'_1}\dots
g^{a_na'_n}(g^{-1})^{b_1b'_1}\dots(g^{-1})^{b_mb'_m}=
\nonumber\\
& &=\left.\frac{\delta}{\delta K^{a_1a'_1}(b)}\dots
\frac{\delta}{\delta K^{a_na'_n}(b)}
\frac{\delta}{\delta K^{b_1b'_1}(-b)}
\dots\frac{\delta}{\delta K^{b_mb'_m}(-b)}D(K(b),K(-b))\right|_{K=0}
\eea
The structure of the expansion is revealed considering the integrals over the
group elements on the path $\om$. A general discussion of
integration over SU(N) group elements is given by Creutz \cite{CR}. For our
purposes
it is useful to know the following two results (together with the Haar
measure normalization $\int dg=1$):
\bea
\int dg g^{aa'}&=&0\\
\int dg g^{aa'}(g^{-1})^{bb'}&=&\frac{1}{N}\delta_{ab'}\delta_{a'b}
\eea
We consider a closed path $\om$ which forms a loop from a point $x\in\Lambda$
to a point
$y\in \Lambda$ and returns to x
\bea
\om&=&\om^+\cup \om^-\nonumber\\
\om^+&:&x\longmapsto y\nonumber\\
\om^-&:&y\longmapsto x
\eea
(the path can have any other closed form, it can be, for instance,
the union of two disconnected closed paths: a loop from x to x and another loop
from y to y).
The graphs with non trivial group integral are given by the following
algorithm:
\begin{punto}{The path} We draw the path $\om$ on the lattice $\Lambda$.
We choose a direction of the path. Each bond $b\in\om$ defines a link between
two points. The direction of this link is defined by the direction of the path.
\end{punto}
\begin{punto}{Plaquettes addition} In order to have
a non vanishing group integral over a hypercolor singlet polynomial
of group matrices we have to add plaquettes in the following way.\\
A plaquette is bounded by four directed
links. For each given path $\om$
we add plaquettes on the lattice $\Lambda$ in such a way that there are
2n (n is some integer number) or zero links for each bond $b\in\Lambda$
and if there are 2n links at a bond b, half of the
links are directed along $+b$ and the other half along $-b$.
\end{punto}
\begin{punto}{Vertices and bonds} We consider only the bonds with 2n links.
This set of bonds is called the {\em bond set}
and the respective set of start and end points of
these bonds is called {\em vertex set}.
The bond set forms a {\em general graph}.
\end{punto}
\begin{punto}{Connected graph}
We call {\em graph} $G(\om)$ a connected general graph
constructed from a
path $\om$ with the previous three steps of this algorithm.
The graph $G(\om)$ is not unique, but the group integral of two
different $G(\om)$ of the same path $\om$ and with the same number of
vertices is unique.
\end{punto}
\begin{punto}{Group integral} We denote with
$\Omega_{ab'\,a'b}(G(\om))$ the group
integral of the path $\om$ for graph $G(\om)$.
The hypercolor indices aa' and bb' indicate the hypercolor indices at the
begin (a,b) and at the end (a',b') of the paths $\om^+$ and $\om^-$,
respectively.
\begin{punto}{Graph of type I}
We call {\em graphs of type I} a graph\footnote{See Fig. 1a}
constructed from a path $\om$ with the first three steps of this algorithm
with n=1.
All graphs $G(\om)$ of type I of the same path $\om$
and with the same number of plaquettes have
the same number V of vertices and B of bonds. The group integral for a
graph of type I yields a very simple result:
\be
\Omega_{ab'\,a'b}(G(\om))=N^{(V-B-1)}\delta_{ab'}\delta_{a'b}
\ee
A term N arises from
contracting the hypercolor indices of eq. (34) at each vertex and a
term 1/N arises from each bond with two links (see eq. (34)).
\end{punto}
\begin{punto}{Graph of type II} We call
{\em graph of type II} a graph which is not of type I\footnote{See Fig. 1b}.
The group integral for a graph of type II has not a simple form and a
general result like for the previous case. The
group integral has to be evaluated separately following the methods explained
in \cite{CR}.
\end{punto}
\end{punto}

\subsection{Strong coupling Feynman diagram}
A strong coupling Feynman {\em diagram} $D(G,\om)$
is formed by a path $\om$ and {\em one} of its
graphs $G(\om)$. The order of a term in the strong coupling expansion
which is characterized by a diagram $D(G,\om)$
is given by the number P of plaquettes of the graph $G(\om)$
and by the length of the path $|\om|$.
A diagram will give a contribution to the Green's function proportional to
$(-\beta)^P\times k^{|\om|}$. The amplitude $A(D(G,\om))$
of the diagram is given by:
\be
A(D(G,\om))=(-1)^L\,\Omega_{ab'\,a'b}(G(\om))\,
\Sigma_{aa'\,bb'}(\om)\,(-\beta)^Pk^{|\om|}
\ee
where L is the number of internally closed loops formed by the path
and $\Sigma(\om)$
is the trace term arising from the trace over the matrices of spin and
isospin. The trace over the hypercolor is obtained by contracting
the indices aa' and
bb'. This last trace term $\Sigma$ is different from the analogous term that
one expects from a strong coupling expansion with Dirac fermions and its
evaluation requires care. Therefore we will discuss it in a separate
subsection.

\subsection{The trace term}
In this subsection we consider only two-point Green's functions of the form:
$$\langle \fee(x)M\fee(x)\fee(y)M\fee(y)\rangle$$
The normalized Green's function is
$$\langle \fee(x)M\fee(x)\fee(y)M\fee(y)\rangle=\frac{1}{Z}\int
[d\fee][dU]
\fee(x)M\fee(x)\fee(y)M\fee(y)exp\{-S(\fee,U)\}$$
When we consider only the
fermion integration, the normalized fermion expectation value will be
$$\langle \fee(x)M\fee(x)\fee(y)M\fee(y)\rangle(U)^{fermion}=\frac{1}{Z}\int
[d\fee]
\fee(x)M\fee(x)\fee(y)M\fee(y)exp\{-S(\fee,U)\}$$
This can be written as a sum over paths forming
fermion loops connecting x and y. There are two possibilities:
\begin{punto}{Isomultiplet} If the matrix $M$ contains some isospin matrix
$T^I$
there is no contribution from separate loops at x and y, since $T^I$
is traceless. Only
loops connecting x with y are allowed.
\end{punto}
\begin{punto}{Isosinglet} If the matrix $M$ does not contain a $T^I$ term the
contributions from separate loops at x and y are allowed.
If two loops are connected by plaquettes then they
correspond to a chiral anomaly contributions.
If the two loops are not connected by plaquettes (i.e. they do not form a
connected diagram) then the truncated Green's function has to be defined as
follows\footnote{Candidates here are the CP eigenstates $S_\pm=S\pm\bar S$.
However,
one can easily show
that $\langle S_-\rangle=0$ (the contribution of a
graph to $\langle S_-\rangle$ cancels with the contribution of the same graph
with all the space direction reversed) which avoids spontaneous CP
violation and $\langle S_+\rangle\neq 0$ due to a spontaneous breaking of the
U(1) fermion number transformation.
The vacuum expectation value of the vector fields is zero, because
a spontaneous breaking of the Lorentz invariance should not occur.}
\bea
& &\langle \fee(x)M\fee(x)\fee(y)M\fee(y)\rangle^{truncated}=\nonumber\\
& &=\langle \fee(x)M\fee(x)\fee(y)M\fee(y)\rangle
-\langle \fee(x)M\fee(x)\rangle\langle\fee(y)M\fee(y)\rangle
\eea
\end{punto}
For the moment we concentrate ourselves on the
non-anomaly terms. We will discuss the anomaly later in section 5 .
We have defined the closed loop path $\om$ from a point x to a point y as
the union of two distinct paths $\om^+$ and $\om^-$. Each of these two paths
is the union of the $|\om^\pm|$ bonds which compose them
\bea
\om^+&=&\bigcup_{j=1}^{|\om^+|}\langle x_{j-1},x_j\rangle\nonumber\\
\om^-&=&\bigcup_{j=1}^{|\om^-|}\langle y_{j-1},y_j\rangle
\eea
where $x_0=x$, $x_{|\om^+|}=y$; $y_0=y$ and $y_{|\om^-|}=x$.\\
We use the shorthand
$\delta\lambda(x)\equiv \frac{\delta}{\delta \lambda(x)}$
for the differentiation of the fermion
sources.
The contribution of the fermion along the path is expressed by the derivative
of the fermionic sources at each lattice point on the path. The kinetic term,
proportional to the hopping parameter k, yields terms of the following form
\bea
\Theta_{aa'}(\om^+)\tilde Z&=&\left.\delta\lambda^T(x)
\left(\begin{array}{cc}
Q&0\\0&1\end{array}\right)
\Gamma(\om^+)\times\delta_{aa'}\times
\left(\begin{array}{cc}
1&0\\0&Q\end{array}\right)
\delta\lambda(y)\,\tilde Z\right|_{\lambda(z)=0\, for x\neq z\neq y}\nonumber\\
\Theta_{bb'}(\om^-)\tilde Z&=&\left.
\delta\lambda^T(y)
\left(\begin{array}{cc}
Q&0\\0&1\end{array}\right)
\Gamma(\om^-)\times\delta_{bb'}\times
\left(\begin{array}{cc}
1&0\\0&Q\end{array}\right)
\delta\lambda(x)\,\tilde Z\right|_{\lambda(z)=0\, for x\neq z\neq y}\nonumber\\
\eea
where
\bea
\Theta_{aa'}(\om^+)&=&\delta_{aa'}\;\prod_{j=1}^{|\om^+|}\delta\lambda^T(x_{j-1})
\left(\begin{array}{cc}
Q&0\\0&1\end{array}\right)
\Gamma(\langle x_{j-1},x_j\rangle)
\left(\begin{array}{cc}
1&0\\0&Q\end{array}\right)
\delta\lambda(x_j)\nonumber\\
\Theta_{bb'}(\om^-)&=&\delta_{bb'}\;\prod_{j=1}^{|\om^-|}\delta\lambda^T(y_{j-1})
\left(\begin{array}{cc}
Q&0\\0&1\end{array}\right)
\Gamma(\langle y_{j-1},y_j\rangle)
\left(\begin{array}{cc}
1&0\\0&Q\end{array}\right)
\delta\lambda(y_j)
\eea
When we differentiate the
exponential of the quadratic form $\lambda^TA^{-1}\lambda$ in $\tilde Z$
twice with respect to $\lambda$ we have
\bea
& &\delta\lambda_{\tilde\al,a}^A(x)\delta\lambda_{\tilde\beta,b}^B(x)
\times exp{\{-\lambda^T(x)\frac{A^{-1}}{2}\lambda(x)\}}=\nonumber\\
& &=
\left(
A^{-1}_{AB,\tilde\al\tilde\beta,ab}
+(A^{-1}\lambda(x))
_{A,\tilde\al,a}(A^{-1}\lambda(x))_{B,\tilde\beta,b}\right)
\times exp\{-\lambda^T(x)\frac{A^{-1}}{2}\lambda(x)\}
\eea
where
\be
A_{AB,\tilde\al\tilde\beta,ab}=\left(\begin{array}{cc}
Q&0\\0&Q\end{array}\right)
\ee
denotes the {\em totally antisymmetric} matrix in isospin, spin and hypercolor
space. Eq. (40) is obtained using eq. (42) in (41)\footnote{All
terms involving the second term in the bracket on the right hand
side of eq. (42)
will vanish because
of the backtracking (one obtains gamma matrix multiplications
of the form $\Gamma(b)\Gamma(-b)=0$) or because they are proportional to
a $\lambda(x_j)$ which will be set to zero in eq. (40).}.
The external source derivative
$\frac{\delta}{\delta K_{cc'}(\langle x_{j-1},x_j\rangle)}$
associated to the gauge field can be viewed as an identity matrix
$\delta_{cc'}$ because of eq. (34) and the algorithm of sect. 3.2. The
remaining
terms $\delta_{aa'}$ and $\delta_{bb'}$ arise from contracting all the
hypercolor
indices of the $\delta_{cc'}$ for each bond of the path.
The remaining $\delta\lambda(x)$ and $\delta \lambda(y)$ in eq. (40)
will contract their indices with the indices of the derivatives
$\delta\lambda(x)$ and $\delta\lambda(y)$ arising from the propagator
$\langle \fee(x)M\fee(x)\fee(y)M\fee(y)\rangle$ when we substitute the
fermion fields $\fee$ with the derivative $\delta\lambda$. Applying these
derivatives to the quadratic form in $\tilde Z$ we obtain
\be
\Sigma_{aa'\,bb'}(\om)=\left.Tr
\left\{\delta\lambda(x)M
\delta\lambda(x)\times\Theta_{aa'}(\om^+)\times
\delta\lambda(y)M\delta\lambda(y)\times\Theta_{bb'}(\om^-)\tilde
Z\right\}\right|_{\lambda=0}
\ee
Using eq. (40) in (44) we obtain
\bea
& &\Sigma_{aa'\,bb'}(\om)=\delta_{aa'}\delta_{bb'}\times\nonumber\\
& &\times Tr
\left\{\delta\lambda(x)M\delta\lambda(x)\times
\delta\lambda(x)\left(\begin{array}{cc}
Q&0\\0&1\end{array}\right)
\Gamma(\om^+)
\left(\begin{array}{cc}
1&0\\0&Q\end{array}\right)
\delta\lambda(y)\times\right.\nonumber\\
& &\left.\left.\times
\delta\lambda(y)M\delta\lambda(y)\times
\delta\lambda(y)\left(\begin{array}{cc}
Q&0\\0&1\end{array}\right)
\Gamma(\om^-)
\left(\begin{array}{cc}
1&0\\0&Q\end{array}\right)
\delta\lambda(x)\times\tilde Z\right\}\right|_{\lambda=0}
\eea
The result one obtains from eq. (45) may be very different from the analogous
result for Dirac fermions. The difference is due to the
fact that in the Dirac case there are two independent fermionic external
sources
(say $\lambda$ and $\bar\lambda$) while in our case there is
only one. In the Dirac
analogue to eq. (45) there are two $\delta\lambda$ and two
$\delta\bar\lambda$, and the contraction of the indices is
evident. In our case we have four $\delta\lambda$ and also four times the same
functional derivative which leads to a more tricky contraction of the
indices.\\
In eq. (45)  one can contract the
indices of the derivatives $\delta\lambda(x)$ and $\delta\lambda(y)$ in many
different ways, because each $\delta\lambda$ appears four times.
The proper contraction of the $\lambda$ using an explicit notation
leads to a sum of four terms
\bea
& &\Sigma_{aa'\,bb'}(\om)=\delta_{aa'}\delta_{bb'}\times\nonumber\\
& &\times Tr\left\{\Gamma(\omega^+)\tilde M\Gamma(\omega^-)\tilde M
-\Gamma(\omega^+)A\tilde M^TA\Gamma(\omega^-)\tilde M\right. \nonumber\\
& &\left. +\Gamma(\omega^+)A\tilde M^TA\Gamma(\omega^-)A\tilde M^TA
-\Gamma(\omega^+)\tilde M\Gamma(\omega^-)A\tilde M^TA\right\}
\eea
where
\be
\tilde M=\left(
\begin{array}{cc}Q&0\\0&1\end{array}\right)M
\left(\begin{array}{cc}1&0\\0&Q\end{array}\right)
\ee
\section{CP eigenstates}
In weak interactions CP is a good quantum number but not C and P separately.
Therefore, we perform a classification of the composite operators (1-3)
according to the CP eigenvalues.
The CP transformation of the Weyl spinor $F$ is defined by
$$
F^{CP}(x)= i\tau_2i\si_2F^{\dagger}(x_P)
$$
(where $x_P=(t,-\vec x)$) which fixes the phase.
In the compact notation we can rewrite the CP
transformation in the following form:
\bea
\psi^{CP}(x)&=&T_2\gamma_0\psi(x_P)\nonumber\\
\bar\psi^{CP}(x)&=&\bar\psi(x_P)\gamma_0T_2
\eea
We can build CP odd and CP even combinations\footnote{Lorentz
scalars $S$, vectors $V^\mu$ and tensors $T^{\mu\nu}$ are CP {\bf even} if
$CP\,S=S$, $CP\,V^\mu=V_\mu$ and $CP\,T^{\mu\nu}=T_{\mu\nu}$ and are
CP {\bf odd} if
$CP\,S=-S$, $CP\,V^\mu=-V_\mu$ and $CP\,T^{\mu\nu}=-T_{\mu\nu}$, respectively.}
of the composite fields (1),
(2) and (3). These eigenstates can be expressed in a very compact form with the
definitions (19).
\begin{punto}{Scalar CP eigenstates} The scalar CP {\bf even} combination of
scalars is defined by
\be
S_+(x)=\bar\psi(x)\psi(x)
\ee
The pseudoscalar CP {\bf odd} combination is defined by
\be
S_-(x)=i\bar\psi(x)\gamma_5\psi(x)
\ee
\end{punto}
\begin{punto}{Vector CP eigenstates} The vector isosinglet
CP {\bf even} combination is defined by\footnote{Notice that
using eq. (10) we obtain
$$V^\mu= F^\dagger(x)\si^\mu F(x)-F^T(x)Q\tilde\si^\mu QF^{\dagger}(x)=
F^\dagger(x)\si^\mu F(x)+F^\dagger(x)\si^\mu F(x)=2F^\dagger(x)\si^\mu
F(x)$$}
\be
V^{\mu}(x)=\frac{i}{2}\bar\psi(x)\gamma^\mu\gamma_5\psi(x)
\ee
The isotriplet vector states is defined by
\be
V^{\mu I}(x) =\frac{1}{2}\bar\psi(x)\gamma^\mu T^I\psi(x)
\ee
This state is CP {\bf even}
for I=1,3 and CP {\bf odd} for I=2.
\footnote{Notice that the charged and
neutral weak bosons are defined by
$$
\left(\begin{array}{c}W^+\\W_3\\W^-\end{array}\right)=
\left(\begin{array}{c}W_1+iW_2\\W_3\\W_1-iW_2\end{array}\right)
$$
The CP properties are:
$CPW_3(x)=W_3(x_P)$ and $CPW^+(x)=W^-(x_P)$.}
\end{punto}
\begin{punto}{Tensor CP eigenstates} The tensors are defined as
follows:
\be
\tilde B^{\mu\nu I}=\bar\psi(x)\si^{\mu\nu}T^I\psi(x)
\ee
and its dual
\be
\tilde B^{\mu\nu I}_\ast=\epsilon^{\mu\nu\sigma\rho}\tilde B_{\sigma\rho}^I
\ee
where $\si^{\mu\nu}=\frac{1}{2}[\gamma^\mu,\gamma^\nu]$.
$\tilde B^{\mu\nu I}$ has CP properties opposite to the vector isotriplet $V^I$
ones. Its dual $\tilde B^{\mu\nu I}_\ast$ has CP properties opposite to
$\tilde B^{\mu\nu I}$.
\end{punto}
The propagators of the CP eigenstates going
from the point x to the point y along some path $\om$ can be evaluated with
the help of eq. (37) and choosing the
appropriate representation of the matrix $M$ in eq. (46).
In the CP formalism we can write a propagator in the form
\be
\langle(\fee(x)M\fee(x))^\dagger\fee(y)M\fee(y)\rangle=
\langle(\bar\psi(x)\tilde M\psi(x))^\dagger\bar\psi(y)\tilde M\psi(y)\rangle
\ee
Inserting the matrices $\tilde M$ relative to the CP eigenstates
in the eq. (37) and (46) and using eq. (9-11) the result takes a simple form
\bea
& &\langle(\bar\psi(x)\tilde M\psi(x))^\dagger\bar\psi(y)\tilde
M\psi(y)\rangle=\sum_{\omega:x\mapsto y\mapsto x}\,\,
\sum_{G(\omega)}A(D(G,\omega))
=\nonumber\\
& &=\sum_{\omega:x\mapsto y\mapsto x}\,\,\sum_{G(\omega)}
C_{M}\times Tr(\tilde M^\dagger\Gamma(\omega^+)\tilde M\Gamma(\omega^-))
\times(-1)^L\Omega(G(\omega))(-\beta)^Pk^{|\omega|}
\eea
where $C_M$ is some constant and
$\Omega(G(\omega))=\Omega_{ab'a'b}(G(\omega))\delta_{aa'}\delta_{bb'}$. Notice
that for a graph of type I we obtain $\Omega(G)=N^{(V-B)}$ (see eq. (36)).

\section{Strong coupling expansion masses}
In order to compute the mass of the CP eigenstates we will
consider the static propagators of these states.
\be
G^{\tilde M}_{static}(t)=\sum_{\vec{x}}
\langle(\bar\psi((0,\vec{x})){\tilde M}\psi((0,\vec{x})))^\dagger
\bar\psi((t,\vec{x})){\tilde M}\psi((t,\vec{x}))\rangle
\ee
First
we assume that $x_0=0$ and
$y_0=t$ and $\vec{x}=\vec{y}$ (summed over all $\vec{x}$)
to interpret eq. (56) as a static propagator.
The lowest order diagram representing a static propagator is a double fermion
line in the time direction as shown in Fig. 2a. Since there is a non-zero
probability
for a transition from the static state to a more complicated state, the full
propagator is given by a sequence of excited states connected by static
propagators as shown in Fig. 2b.
Each diagram $D(G(\om),\om)$ of a general path $\om$ can be viewed as a
space-time process contributing to the excitation of the static propagation.
These excitations renormalize the mass of the static propagator.

\subsection{The unrenormalized static propagator}
We denote with $G^{\tilde M} (t)$
a static propagator.
For the CP eigenstates we will consider the operators given by the following
set
of ${\tilde M}$'s.\footnote{Because of the relations $\pa_\mu V^\mu=C\times
S_-$
(when the anomaly is neglected)
and $\pa_\mu \tilde B^{\mu\nu I}=C'\times V^{\nu I}$
(where $C$ and $C'$ are some constants) and because of the Fourier
transformation with zero momenta $\vec{p}$ in eq. (57)
the static propagator mixes $S_-$ with $V^0$ and $V^{kI}$ with
$\tilde B^{okI}$.}\\[0.2cm]
\begin{center}
\begin{tabular}{|c|l|}\hline
${\tilde M}$ & flavor \\\hline\hline
$\gamma_5+\gamma_5\gamma_0$ & $\tilde S_-$\\\hline
$\gamma_kT^I+\sigma^{0k}T^I$ & $\tilde V^{kI}$\\\hline
$\gamma_5\gamma_k$ & $V^k$\\\hline
$\sigma^{kj}T^I$ & $\tilde B^{kjI}$\\\hline
1 & $S_+$\\\hline
\end{tabular}\\[0.3cm]
\end{center}
We will use the shorthands
$$\tilde S_-,\,\tilde V^{kI},\,V^{k},\,\tilde B^{kj I},\,S_+$$
to denote the flavor of the bound states and its
corresponding matrix $\tilde M$.
The non renormalized mass of a ${\tilde M}$ state can be found from the
property
of the unrenormalized static propagator (for $t\rightarrow \infty$)
\be
G_0^{{\tilde M}}(t)=C_{\tilde M} \times exp(-m_0t)
\ee
where $C_{\tilde M}$ is a constant.
The path $\om_0$ denotes the path of Fig. 2a. The diagram of this path contains
no plaquettes. Computing the amplitude (37) of this
diagram we have to evaluate the corresponding $\Gamma$ matrices\footnote{For
the notation see eq. (14). We have fixed r=1 which
correspond to the case r=1 in QCD.}
\bea
\Gamma(\om_0^+)=(\Gamma^0)^t=2^{(t-1)}\Gamma^0\nonumber\\
\Gamma(\om_0^-)=(\bar\Gamma^0)^t=2^{(t-1)}\bar\Gamma^0
\eea
The group integral of this diagram is $\Omega(\om)=2$
and so we find from eq. (56) that
\be
G_0^{{\tilde M}}(t)= C_{\tilde M}\times (4k^2)^t
\ee
The static unrenormalized propagator $G_0^{\tilde M}(t)$ is
different from zero only for two CP eigenstates:
\be
{\tilde M}=\tilde S_-,\,\tilde V^{kI}
\ee
{}From eq. (56) we find that for these bound
states the non renormalized mass is
\be
m_0=-\log(4k^2)
\ee
The static propagator of the other CP eigenstates is zero.
Their mass is
$m_0=\infty$ and remain very large even
after renormalization.
This result is a pathologic
default of the strong coupling expansion. It does not describes any
physics. The only information that one can obtain after renormalization is
the spectrum of the states for which $G_0(t)\neq 0$. In the next
subsections we will compare the masses of these two states.

\subsection{Renormalization of the static propagator}
Having computed the static propagator, we will now compute the renormalization
produced by intermediate exited states. Fig. 3 shows us such an excitation,
with initial point z and final point w, where the static
unrenormalized propagator arrives and
departs. For fixed z and w, we will sum over all intermediate exited states to
obtain a total weight for the event. We denote this weight by $D^{\tilde
M}(z,w)$. The
full propagator takes the form \cite{FR}
\bea
G^{{\tilde M}}(x,y)&=&\sum_{z_i,w_i(i=1\dots
n)}G_0^{{\tilde M}}(x,z_1)D^{\tilde M}(z_1,w_1)G_0^{{\tilde
M}}(w_1,z_2)D^{\tilde M}(z_2,w_2)\times\nonumber\\
& &\times\dots \times D^{\tilde M}(z_n,w_n)G_0^{{\tilde M}}(w_n,y)
\eea
where the points $z_i$ and $w_i$ are required to lie between x and y and to be
ordered so that each $w_i$ is "later" than $z_i$, which is itself "later" than
$w_{i-1}$. This last eq. (63) represents the picture in Fig. 2b.\\
In order to find the renormalized mass of the propagator, we consider
the full static propagator
\be
G^{{\tilde M}}(t)=\sum_{y_0-x_0=t}G(x,y)
\ee
Similarly we define\footnote{Notice that
$(G_0^{\tilde M}(t))^{-1}=\frac{1}{C^{\tilde M}} e^{m_0t}=\frac{1}{C^{\tilde
M}} (4k^2)^{-t}$}
\be
D^{\tilde M}(t)=(G_0^{\tilde M}(t))^{-1}\sum_{y_0-x_0=t}D^{\tilde M}(x,y)
\ee
and then (63) gives
\bea
G^{{\tilde M}}(t)&=&e^{-m_0t}
\sum_{s_i,t_i(i=1\dots n)}D^{\tilde M}(t_1-s_1)\dots D^{\tilde M}(t_n-s_n)=
\nonumber\\
&\simeq&e^{-m_0t}\,e^{p_{{\tilde M}}t}
\eea
where $t_i$ and $s_i$ are the corresponding time coordinates of $w_i$ and
$z_i$. So the renormalized mass $m_{{\tilde M}}$ is
\be
m_{{\tilde M}}=m_0-p_{{\tilde M}}
\ee
The leading order of $p_{\tilde M}$ can just be written in the form
\be
p_{{\tilde M}}=
\sum_{w=(t'\geq 0,\vec w)}(G_0^{\tilde M}(t'))^{-1}D^{\tilde M}(0,w)=
\sum_{w=(t'\geq 0,\vec w)}(4k^2)^{-t'}D^{\tilde M}(0,w)
\ee
To proof this last equation we have to insert it into the last term of (66) and
to expand the exponential function containing $p_{\tilde M}$, then we have to
compare the result with the first term of (66). The excitation term
$D^{\tilde M}(z,w)$ is defined starting from eq. (37) in the following way:
\be
D^{\tilde M}(z,w)=\sum_{\om:z\mapsto w\mapsto z}\,\,\sum_{G(\om)}
\frac{A(D(G,\om))}{C^{\tilde M}}
\ee
where the sum is over all closed paths $\om$ from z to w and return
and over all graphs $G(\om)$. $A(D(G,\om))$ denotes the
amplitude (37) of an intermediate excitation diagram.

\subsection{Results}
In order to calculate the renormalized mass of the two states $\tilde V^{kI}$
and $\tilde S_-$ to next leading order
we have collected all the excitation diagram for all $t'\geq 0$ and $\vec{w}$
in eq. (68) corresponding to the $k^4$ and $\beta^4$ perturbation order.
Our results are valid in the asymptotic sense in a
neighborhood of the origin in the $(k,\beta)$
plane.
\begin{punto}{Vector isotriplet mass} First we write the vector isotriplet
mass $m_{\tilde V^{kI}}$ up to
fourth order in $\beta$ and k.
Our result is
\bea
m_{\tilde V^{kI}}&=&-log(4k^2)
-3k^2+\frac{3}{2}k^2\beta-\frac{3}{4}k^2\beta^2+\frac{7}{8}
k^2\beta^3-\frac{5}{16}k^2\beta^4\nonumber\\
& &-16k^4+15k^4\beta-\frac{107}{4}k^4\beta^2+14k^4\beta^3-\frac{272}{3}k^4
\beta^4
\eea
\end{punto}
\begin{punto}{Pseudoscalar mass} We calculate the pseudoscalar mass
neglecting the anomaly terms, since they arise only at order $k^{14}$ and are
generally severely suppressed in strong coupling expansions \cite{AM}.
These terms will be discussed later.
In this calculation the pseudoscalar mass comes out to be less than
the vector mass. This is not surprising in view of the suppression of the
anomaly.
In the next subsection the consideration of the anomaly terms
will show that the pseudoscalar bound state behaves like the $\eta'$ bound
state
in QCD and in particular that it acquires a positive contribution to its mass
from the anomaly terms. Our result is
\bea
m_{\tilde S_-}&=&-log(4k^2)-3k^2+\frac{3}{2}k^2\beta-\frac{3}{4}k^2\beta^2+
\frac{7}{8}k^2\beta^3-\frac{5}{16}k^2\beta^4
\nonumber\\
& &-36k^4+12k^4\beta-\frac{153}{4}k^4\beta^2+\frac{65}{4}k^4\beta^3-
\frac{237}{2}k^4\beta^4
\eea
\end{punto}

\subsection{The anomaly terms}
Here we will calculate the leading order anomaly
contribution to the mass
splitting $\Delta m_{\tilde S_-}^{anomaly}=m_{\tilde S_-}^{tot}-m_{\tilde S_-}$
where $m_{\tilde S_-}^{tot}$ denotes the pseudoscalar mass when the anomalies
are turned
on. We use the methods derived in section 3 to determine the contributions of
the anomaly terms.
The mass splitting is given by excitations which are characterized by two
disconnected closed loops:
\bea
\om^{anomaly}&=&\om^1\cup\om^2\nonumber\\
\om^1&:&x\longmapsto x\nonumber\\
\om^2&:&y\longmapsto y\nonumber
\eea
The diagram which corresponds to these two paths can be found by the
algorithm of section 3.2. The amplitude of this diagram is given by the formula
(37). The trace term can be evaluated by modifying eq. (45) to the new
paths type. The result is
\bea
& &\Sigma(\om^{anomaly})=\nonumber\\
& &=Tr\left\{\delta\lambda(x)M_1\delta\lambda(x)\times
\delta\lambda(x)\left(\begin{array}{cc}
Q&0\\0&1\end{array}\right)
\Gamma(\om^+)
\left(\begin{array}{cc}
1&0\\0&Q\end{array}\right)
\delta\lambda(x)\times\right.\nonumber\\
& &\times \left.\left.
\delta\lambda(y)M_2\delta\lambda(y)\times
\delta\lambda(y)\left(\begin{array}{cc}
Q&0\\0&1\end{array}\right)
\Gamma(\om^-)
\left(\begin{array}{cc}
1&0\\0&Q\end{array}\right)
\delta\lambda(y)\times\tilde Z\right\}\right|_{\lambda=0}
\eea
For the pseudoscalar bound state $\tilde S_-$ the result takes the form
\be
\Sigma_{\tilde S_-}(\om^{anomaly})=4\left[Tr\left(\gamma_5\Gamma^0
\Gamma(\om^1)\right)\times
Tr\left(\gamma_5\Gamma^0\Gamma(\om^2)\right)\right]
\ee
To find all paths which yield a non trivial contribution to the trace term (75)
one has to find all the loops $\om^i$ (i=1,2) for which
\be
Tr[\gamma_5\Gamma^0\Gamma(\om^i)]\neq 0
\ee
Clearly this last equation is true if $\Gamma(\om^i)$ contains a $\gamma_5$.
This condition requires that $\om^i$ covers bonds in any direction,
since $\gamma_5=\gamma_0\gamma_1\gamma_2\gamma_3$.
The reader can convince
himself that the shortest loops must have at least eight bonds.
The allowed lowest order loops are plotted in Fig. 4. The spin trace (74)
is the same for all these loops, up to a sign:
$Tr[\gamma_5\Gamma^0\Gamma(\om^i)]=\pm 32$.\\
An excitation diagram which contributes to the anomaly mass splitting is
composed by
two of the loops in Fig. 4 (one for each end of the excitation) connected
by plaquettes to make a hypercolor singlet as required by the
algorithm of section 3.2.
The smallest excitations that can be made in this way are superpositions of two
copies of the same loop, going in opposite direction. The lowest order of
$\Delta m_{\tilde S_-}$ is given by the sum of all such diagrams. It is
proportional
to $k^{14}\beta^0$ with a positive coefficient.\\
The leading order contributions to $\Delta m_{\tilde S_-}$ involving  $\beta$
are
proportional to $k^{14}\beta^n$ (n=1,...,4). These terms are generated
by excitations made out of
two of the loops shown in Fig.4, linked by n plaquettes.
It is easy to check
from these excitations that when
$\beta$ is turned on the effect is to increase the mass of the
pseudoscalar bound state by
\be
\Delta m_{\tilde S_-}^{anomaly}
=2^{12}k^{14}\left(405-216\beta+\frac{189}{2}\beta^2
-\frac{189}{8}\beta^3+\frac{189}{32}\beta^4\right)
\ee

\subsection{Conclusion}
We have obtained two classes of particles in this theory.
\begin{punto}{Finite mass} There are two bound states which have finite
mass in the strong coupling expansion:
The pseudoscalar bound state $\tilde S_-$ and the vector isotriplet bound state
$\tilde V^{kI}$.
The pseudoscalar bound state
$\tilde S_-$ turns out to be lighter than the vector isotriplet bound
state in the strong coupling expansion.
It behaves like
the $\eta'$ bound state in QCD and
can acquire a larger mass by the effect of the chiral anomaly.
\end{punto}
\begin{punto}{Heavy mass} The composite operators
$S_+,\tilde B^{ijI}$ and $V^k$
(see sect. 5.1) correspond in the strong coupling expansion to states with
infinite mass. Such a behavior is known in QCD, where for example the axial
vector
meson gets an infinite mass in the strong coupling expansion \cite{KA}. In
fact, in
nature, axial vector mesons are much heavier than the vector mesons.
\end{punto}
The interesting feature of the use of Majorana fermions as basic constituents
is that the spin one isotriplet bound states (having the correct
assignment to be candidates for the weak bosons) can be the lightest bound
state, provided
that the chiral anomaly rises the mass of the pseudoscalar isosinglet
sufficiently.\\
We have used the results of this work as the starting point of
a detalied investigation of the low lying spectrum of
this theory by a lattice Monte Carlo simulation \cite{galli}.
In particular we have numerically
evaluated the contribution of the chiral anomaly to the pseudoscalar isosinglet
mass.\\[1cm]
{\Large {\bf Acknowledgements}}\\[0.5cm]
We would like to thank F.Jegerlehner and
H.Schlereth for discussion and for reading the manuscript.\\

\vspace{1cm}
{\Large { Figure Caption}}\\[0.2cm]
\begin{enumerate}
\item { a)} A possible graph of type I. { b)} A possible graph of type II.
\item { a)} A static unrenormalized propagator. { b)} An excitation of the
static unrenormalized propagator.
\item Close up of an intermediate excitation.
\item The four geometrically distinct loops which contribute to the trace (74).
\end{enumerate}
\end{document}